\documentclass[useAMS,usenatbib,usegraphicx]{mn2e}

\title[ALMA \& Herschel Observations of G29-38]{ALMA and {\em Herschel} Observations of the Prototype Dusty 
and Polluted White Dwarf G29-38}

\author[J. Farihi et al.]{J. Farihi$^{1,2}$\thanks{E-mail: j.farihi@ucl.ac.uk}\thanks{STFC Ernest Rutherford Fellow},
M. C. Wyatt$^2$,
J. S. Greaves$^3$, 
A. Bonsor$^{4,5}$, 
B. Sibthorpe$^6$, 
O. Pani\'c$^2$\\
$^1$Department of Physics and Astronomy, University College London, London WC1E 6BT\\
$^2$Institute of Astronomy, University of Cambridge, Cambridge CB3 0HA\\
$^3$School of Physics and Astronomy, University of St Andrews, St Andrews KY16 9SS\\
$^4$School of Physics, University of Bristol, Bristol BS8 1TL\\
$^5$Institute de Plan\'etiology, University\'e Joseph Fourier, BP 53, 38041, Grenoble Cedex 9, France\\
$^6$SRON Netherlands Institute for Space Research, P.O. Box 800, 9700 AV Groningen, The Netherlands}

\begin{document}

\date{}

\maketitle

\label{firstpage}

\begin{abstract}

ALMA Cycle 0 and {\em Herschel}\,$^1$ PACS observations are reported for the prototype, nearest, and brightest
example of a dusty and polluted white dwarf, G29-38.  These long wavelength programs attempted to detect an outlying, 
parent population of bodies at $1-100$\,AU, from which originates the disrupted planetesimal debris that is observed within 
0.01\,AU and which exhibits $L_{\rm IR}/L_*=0.039$.  No associated emission sources were detected in any of the data 
down to $L_{\rm IR}/L_*\sim10^{-4}$, generally ruling out cold dust masses greater than $10^{24}-10^{25}$\,g for reasonable 
grain sizes and properties in orbital regions corresponding to evolved versions of both asteroid and Kuiper belt analogs.  
Overall, these null detections are consistent with models of long-term collisional evolution in planetesimal disks, and the 
source regions for the disrupted parent bodies at stars like G29-38 may only be salient in exceptional circumstances, such 
as a recent instability.  A larger sample of polluted white dwarfs, targeted with the full ALMA array, has the potential to 
unambiguously identify the parent source(s) of their planetary debris.

\end{abstract}


\section{INTRODUCTION}


\addtocounter{footnote}{1}
\footnotetext{{\em Herschel} is an ESA space observatory with science instruments provided by European-led Principal 
Investigator consortia and with important participation from NASA.}

\subsection{Discovery and Characterization of G29-38}

More than a quarter century has passed since the discovery of infrared excess emission from the nearby white dwarf G29-38 
\citep{zuc87}.  Photometric observations conducted at the NASA Infrared Telescope Facility atop Mauna Kea in the $K$, $L$, 
and $M$ bands revealed flux in excess of that expected for the relatively cool white dwarf.  While a warm dust disk was considered 
a possibility, the infrared excess was initially attributed to a spatially unresolved brown dwarf.  In particular, 1000\,K circumstellar dust 
was considered unlikely due to rapid dissipation from radiation drag.  Prophetically, \citet{zuc87} noted that if material were orbiting 
sufficiently close to achieve such a high temperature, then spectral signatures of accretion should be seen.

Within a few years, and thanks to intense observational and theoretical interest from a diverse set of researchers, evidence 
began to disfavor a substellar companion as the origin of the infrared emission.  First, some of the earliest infrared imaging 
arrays revealed G29-38 to be a point source in all available bandpasses.  Second, near-infrared spectroscopy measured a 
thermal continuum \citep{tok88}, whereas a very cool atmosphere was expected to exhibit absorption features.  Third, the 
detection of optical stellar pulsations echoed in the near-infrared were difficult to reconcile with a brown dwarf secondary 
\citep{pat91,gra90}.  Fourth and finally, significant 10\,$\mu$m emission was detected at G29-38, at a level a few times 
greater than expected for an object with a Jupiter-sized radius, essentially ruling out the brown dwarf companion hypothesis 
\citep{tok90,tel90}.

A decade after the discovery of its infrared excess, the optical and ultraviolet spectroscopic detection of multiple metal species
in the atmosphere of G29-38 \citep{koe97} made it clear that the star was accreting from its immediate environs.  Although white 
dwarfs with metal absorption features have been known for nearly a century (vMa\,2; \citealt{van17}), the source of the heavy 
elements had never been observationally identified and the closely orbiting disk at G29-38 was the smoking gun.  

\subsection{Tidally-Destroyed Planetesimals}

In a seminal paper, \citet{jur03} modeled the observed properties of G29-38 by invoking a tidally-destroyed minor planet (i.e.\ 
large asteroid) that evolves into an opaque, flat ring of dust analogous to the rings of Saturn.  The particles are heated by the 
star, producing an infrared excess, and slowly dragged down onto the stellar surface, polluting its otherwise-pristine atmosphere 
with heavy elements.  The tidally-disrupted asteroid model has seen continued success since its inception, and is considered the 
standard model for metal-enriched white dwarfs.  In the intervening decade, an enormous amount of observational progress has 
occurred (for a detailed review, see \citealt{far11a}), all of which supports the accretion of asteroid-like debris in dynamically active, 
post-main sequence planetary systems \citep{ver13}.

\medskip
\begingroup
\parindent=6pt\everypar={\hangindent=18pt}

1.  Over 30 metal-lined white dwarfs are now known to exhibit $T\sim1000$\,K thermal emission from disks (e.g.\ \citealt{xu12,
bri12,gir12,far12,far10b,far09,jur07,von07,kil06}); their properties are precisely as expected for material contained within the 
Roche limit of the star and feeding the stellar surface \citep{met12,raf11,boc11}.  A fraction of these exhibit metallic, gaseous 
emission \citep{far12,mel12,mel11,gan08,gan07,gan06} that is spatially coincident with the particulate disks \citep{bri12,mel10}.
\medskip

2.  All dusty white dwarfs observed with IRS on {\em Spitzer} exhibit strong silicate emission features consistent with olivines,
\citep{jur09a,rea09,lis08,rea05}, and which are also seen in the infrared spectra of evolved solids associated with planet formation 
\citep{lis08}.  
\medskip

3.  The elemental abundances in disk-polluted white dwarfs are universally depleted in volatile elements (especially carbon and 
hydrogen), and have a refractory rich pattern that broadly mimics the terrestrial material of the inner Solar System \citep{jur14,gan12,
kle10,zuc07}.  All data acquired to date are consistent with rocky parent bodies that formed interior to a snow line \citep{far13,jur12}, 
including evidence for differentiation \citep{jur13,gan12,zuc11,far11b}.
\medskip

4.  The collective properties of all published ($N\sim250$ as of this writing), metal-polluted white dwarfs support the accretion 
of planetary debris for the population at large \citep{zuc10,far10a,zuc03}, including even those stars that do not exhibit detectable 
infrared excesses \citep{roc14}.  These findings imply that at least 30\% \citep{koe14} of cool white dwarfs have the signatures of
large planetesimals, whose size estimated diameters range from $\sim10$ to $\sim1000$\,km \citep{wya14,jur14}.

\endgroup

\begin{figure*}
\includegraphics[width=126mm]{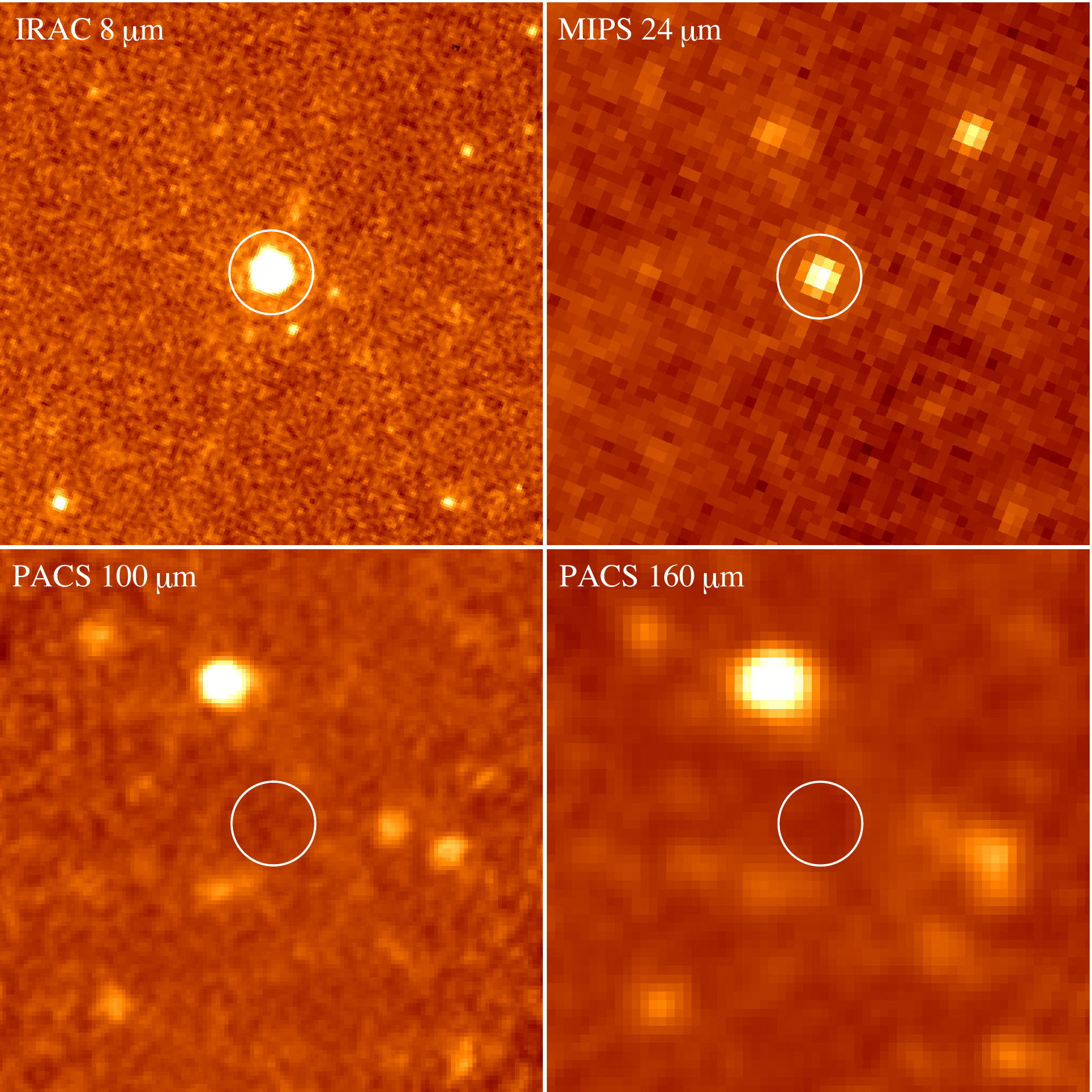}
\caption{Reduced infrared images of G29-38 taken with both {\em Spitzer} and {\em Herschel}.  The images are north up and east 
left and approximately $125''$ on a side.  Accounting for its proper motion of $0\farcs49$\,yr$^{-1}$, the expected location of G29-38 
is marked by a circle in each image, with emission only detected in the {\em Spitzer} data.
\label{fig1}}
\end{figure*}

\begin{table}
\begin{center}
\caption{Multi-wavelength Fluxes and Upper Limits for G29-38\label{tbl1}} 
\begin{tabular}{@{}lcc@{}}
\hline

Source		&$\lambda_{\rm eff}$			&$F_{\nu}$\\
			&($\mu$m)					&(mJy)\\
\hline

{\em GALEX}			&\phantom{000}0.15				&\phantom{$<$ 0}3.4\\
{\em GALEX}			&\phantom{000}0.23				&\phantom{$<$ }10.9\\
$U$					&\phantom{000}0.37				&\phantom{$<$ }17.3\\
$B$					&\phantom{000}0.44				&\phantom{$<$ }21.7\\
$V$					&\phantom{000}0.55				&\phantom{$<$ }22.3\\
$R$					&\phantom{000}0.64				&\phantom{$<$ }18.8\\
$I$					&\phantom{000}0.80				&\phantom{$<$ }15.1\\
$J$					&\phantom{000}1.24				&\phantom{$<$ 0}8.9\\
$H$					&\phantom{000}1.66				&\phantom{$<$ 0}6.0\\
$K_s$				&\phantom{000}2.16				&\phantom{$<$ 0}5.6\\
IRAC				&\phantom{000}3.55				&\phantom{$<$ 0}8.4\\
IRAC				&\phantom{000}4.49				&\phantom{$<$ 0}8.8\\
IRAC				&\phantom{000}5.73				&\phantom{$<$ 0}8.4\\
IRAC				&\phantom{000}7.87				&\phantom{$<$ 0}8.4\\
IRTF					&\phantom{00}10.5\phantom{0}		&\phantom{$<$ }11.1\\
IRS					&\phantom{00}16.0\phantom{0}		&\phantom{$<$ 0}3.7\\
MIPS				&\phantom{00}23.7\phantom{0}		&\phantom{$<$ 0}2.4\\
MIPS				&\phantom{00}71.4\phantom{0}		&\phantom{0}$<$ 1.3\\
PACS				&\phantom{0}100\phantom{.00}		&\phantom{0}$<$ 1.6\\
PACS				&\phantom{0}160\phantom{.00}		&\phantom{0}$<$ 4.4\\
ALMA				&\phantom{0}870\phantom{.00}		&\phantom{00}$<$ 0.17\\
ALMA				&1305\phantom{.00}				&\phantom{00}$<$ 0.21\\

\hline
\end{tabular}
\end{center}

{\em Note}.  Fluxes are {\em GALEX} far- and near-ultraviolet \citep{mar05}, ground-based $UBVRIJHK_sN$ \citep{lan07,skr06,
tok90}, {\em Spitzer} IRAC, IRS 16 and MIPS 24\,$\mu$m photometry \citep{far08,rea05}.  Also listed are $3\sigma$ upper limits 
from MIPS 70\,$\mu$m photometry \citep{jur09b}, and the similar PACS and ALMA limits reported here.

\end{table}

\subsection{Motivation for Long Wavelength Data}

All successful models for disks at white dwarfs invoke tidally-destroyed planetary bodies that originate in a more distant and 
substantially more massive reservoir of planetesimals \citep{fre14,deb12,bon11,jur03,deb02}.  To date, there have been no 
indications of cooler dust associated with an outlying planetesimal population around metal-polluted white dwarfs, but the longest 
wavelength observations conducted for a substantial number of stars are both {\em WISE} 22\,$\mu$m and {\em Spitzer} MIPS 
24\,$\mu$m photometry \citep{hoa13,far09,jur07}.  These data only probe for dust at radii of 1 to a few AU for typical $10\,000-20\,
000$\,K white dwarfs, and thus longer wavelength data are needed to search for populations analogous to the asteroid and Kuiper 
belts of the Solar System.

This paper presents {\em Herschel} \citep{pil10} far-infrared and ALMA submillimeter observations of G29-38 (ZZ\,Psc, WD\,2326$
+$049).  This iconic stellar remnant has a hydrogen-dominated atmosphere of $T_{\rm eff}\approx12\,000$\,K, intrinsic brightness 
$\log(L/L_*)\approx-2.5$, and a cooling age near 380\,Myr \citep{gia12,fon01}.  At 13.6\,pc, G29-38 is the nearest example of a 
disk-polluted white dwarf and also the brightest by an order of magnitude, making it ideal for long wavelength observations sensitive 
to cold dust emission, and where a ring of $20-200$\,AU diameter would span $1\farcs4-14''$ on the sky and be potentially resolved.  

Both sets of observations resulted in null detections, and provide limits on cold dust masses similar to known Kuiper belt objects.  
The observations and data analysis details are presented in \S2, from which are derived sensitivities to fractional dust luminosity,
and similarly to dust masses, as a function of temperature and orbital radius for all existing $\lambda\geq70\,\mu$m observations.
The results are presented in \S3 together with a comparison of these data and sensitivities with known dusty A-type stars, which
represent possible progenitors of the G29-38 system.

\section{OBSERVATIONS AND DATA}

\subsection{{\em Herschel}}

G29-38 was targeted by the {\em Herschel Space Observatory} on 2012 June 6 with the Photodetector Array Camera 
and Spectrometer (PACS; \citealt{pog10}) at 100 and 160\,$\mu$m.  These data are particularly sensitive to 20\,K dust in the 
10\,AU region, where any emission at these wavelengths is expected to be unresolved at the $7''/100$\,$\mu$m diffraction limit 
of a 3.5\,m telescope.  While circumstellar dust orbiting beyond $50-80$\,AU can be spatially resolved around nearby stars with 
{\em Herschel} (e.g.\ \citealt{boo13}), the very low luminosity of a cool white dwarf like G29-38 makes such a detection unlikely
(see \S3.1).

The source was expected to be substantially fainter than the 50\,mJy limit recommended for standard chop-nod observations, and 
thus the mini-scan map mode was used.  These were executed in three pairs of cross-scans, using individual cross-scan angles 
of 70\degr and 110\degr, and each of the six segments repeated 25 times for an on-source time of 1800\,s.  In total, G29-38 was 
observed for 3.0\,hr on source and with 9.4\,hr observatory time.

The PACS data were reduced using the {\em Herschel} Interactive Processing Environment (HIPE) version 7.0.  The data were 
processed using the standard PACS photometer processing steps, and maps were made using the {\sc photProject} task.  The 
data were high-pass filtered in the scan direction with filter width of $66''$ and $102''$ (equivalent to 16 and 25 frames) in the 100 
and 160\,$\mu$m bands respectively.  The fully reduced images are shown in Figure \ref{fig1} alongside {\em Spitzer} imaging
detections in the mid-infrared.

The source was assumed to be point-like and photometry was performed using apertures.  Source apertures with radii of $5''$ and 
$8''$ ($0.7\times$FWHM, which is near optimal for a point source) were used for the 100 and 160\,$\mu$m bands respectively. 
The field background and 1$\sigma$ noise were obtained by taking the median and standard deviation of multiple apertures, with 
radii equal to that of the source apertures, located within $1'$ of the source location respectively.  The measured fluxes ($F_{100,
160\mu{\rm m}}=-0.5,-0.6$\,mJy), and sum in noise apertures ($\sigma_{100,160\mu{\rm m}}=0.7,1.6$\,mJy), have all had aperture 
corrections applied; 1.94 and 1.90 for the 100 and 160\,$\mu$m bands respectively.  The derived upper limits for G29-38 are the 
flux measured in the source apertures $+3\sigma$, and are listed in Table \ref{tbl1}.


\begin{figure}
\centering
\includegraphics[width=64mm]{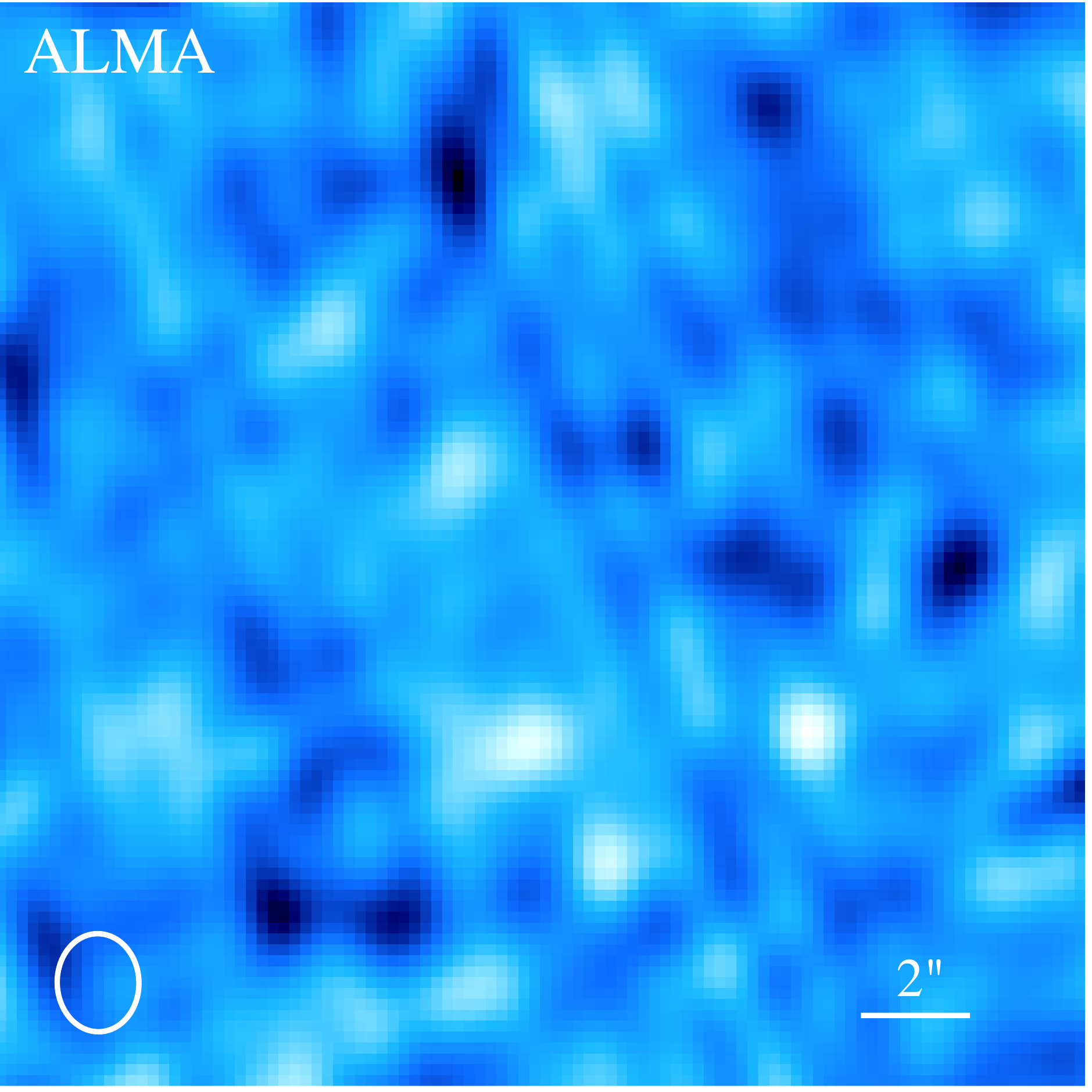}
\caption{Bands 6 and 7, beam-combined ALMA image centered on the expected position of G29-38.  The image is north up and east 
left, with the band 7 beam size and orientation shown in the lower left, and the spatial scale given in the lower right.  No sources are 
detected above $3\sigma\approx0.17$\,mJy. 
\label{fig2}}
\end{figure}

\subsection{ALMA}

G29-38 was also observed with the Atacama Large Millimeter/Submillimeter Array (ALMA) as part of Early Science operations at 
the beginning of Cycle 0 in 2011 November.  Data were acquired over the course of several nights, with two 17\,min observations 
on 18 and 26 November in band 6 (230\,GHz, 1305\,$\mu$m), and five 23 minutes observations on  6, 14, and 16 November in 
band 7 (345\,GHz, 870\,$\mu$m).  During the brief time span of the observations, the proper motion of G29-38 was negligible for 
the concatenation of datasets.  All but two observations were made using 15 antennas, but a single observation in each of band 6 
and 7 was made with 17 and 14 antennas respectively.  The spatial configuration of the array provided baselines covering a range 
from 10 to 150\,m.

Initial calibrations and pointing were done on the bright quasars 3C\,454.3 and 3C\,446, while Neptune was used for absolute flux 
calibration.  The science observations were interleaved with the phase calibrator J2323$-$032.  The observing setup was the most 
sensitive for continuum observations, using the wide-band TDM mode with 128 15625\,MHz wide spectral channels, 2\,MHz 
bandwidth per polarization, and resulting in a full effective bandwidth of 7.5\,MHz.  

\begin{figure}
\centering
\includegraphics[width=84mm]{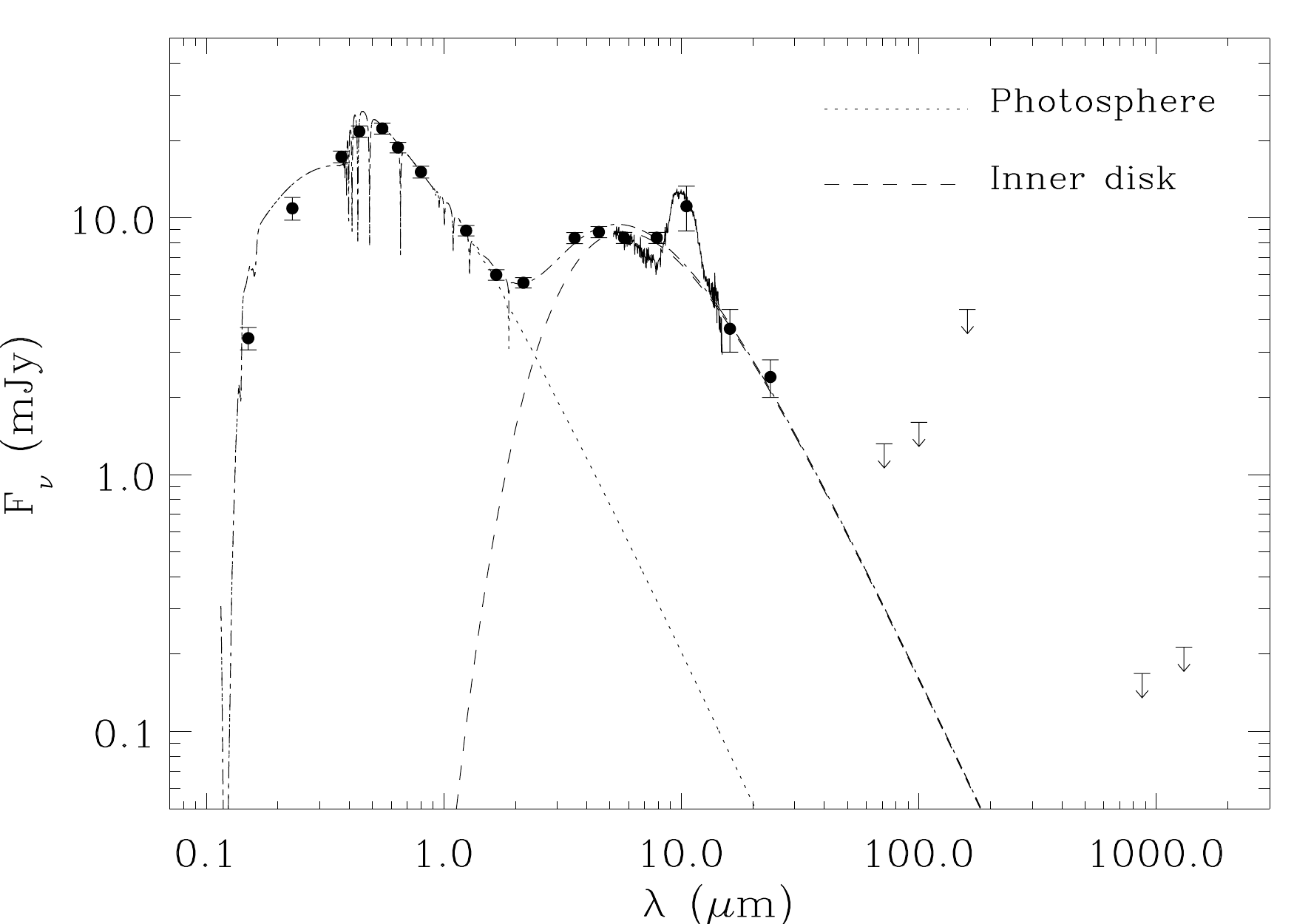}
\caption{Extended spectral energy distribution of G29-38.  Details of the measured fluxes and 3$\sigma$ upper limits are listed in 
Table \ref{tbl1} and discussed in \S2.  The dotted line is a stellar atmosphere model, the dashed line is a face-on, flat and optically
thick ring model ($T_{\rm in}=1250$\,K, $T_{\rm in}=650$\,K) fitted to the warm disk continuum emission, and the solid line the 
measured strong silicate feature in the {\em Spitzer} IRS SL spectrum \citep{rea09}.
\label{fig3}}
\end{figure}

The data were processed by the ALMA pipeline and the achieved RMS values were 0.071\,mJy/beam in band 6 and 
0.056\,mJy/beam in band 7; these were adopted as the $1\sigma$ noise values.  No sources were detected in either band, and 
upper limits for unresolved emission from the science target were taken to be $3\sigma$ and are listed in Table \ref{tbl1}.  Using the 
Common Astronomy Software Applications package, a band 6 and 7, beam-combined image was created to increase the chance of 
source detection using the increased signal and spectral information.  No emission from the science target is evident in these merged 
down to $3\sigma\approx0.17$\,mJy. The beam-combined image is displayed in Figure \ref{fig2}, and is overplotted with the band 7 
beam size ($1\farcs5\times1\farcs8$ at position angle 1.63\degr).

\section{ANALYSIS AND DISCUSSION}

\begin{figure*}
\centerline{\includegraphics[width=188mm]{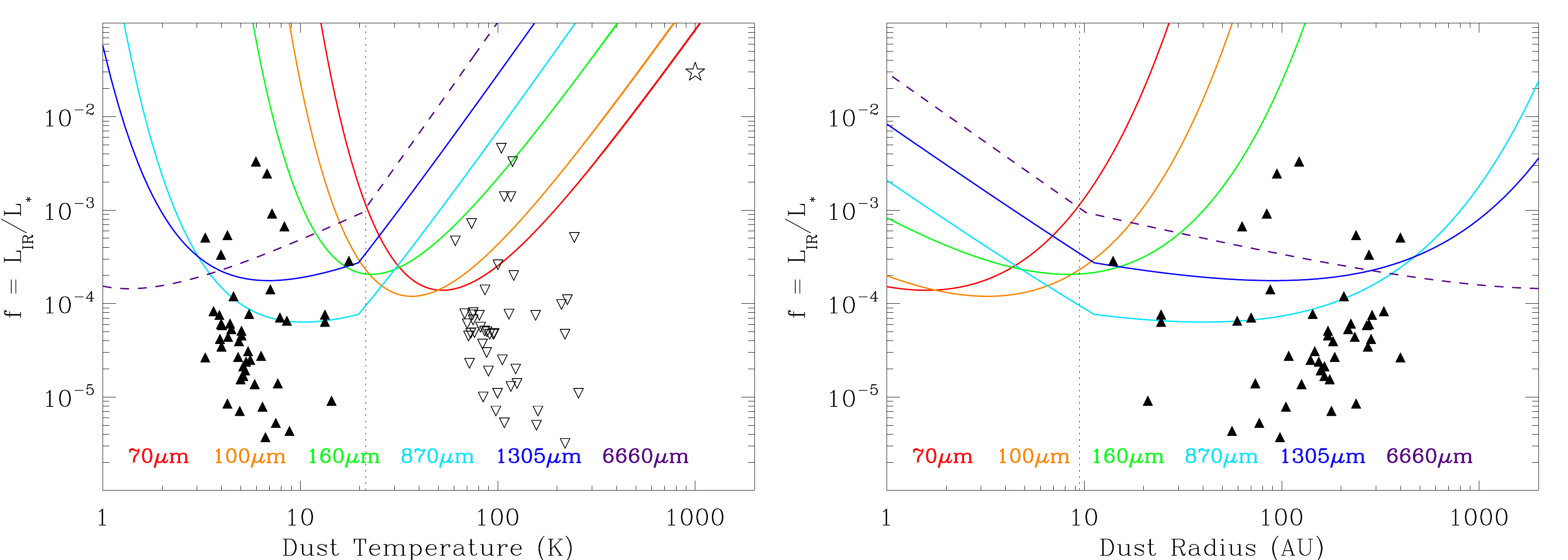}}
\caption{Upper limit sensitivity curves for all long wavelength non-detections of G29-38 as a function of dust fractional luminosity, 
temperature, and orbital radius; only the area above each curve was detectable in individual observations. Limits with {\em Spitzer} 
MIPS 70\,$\mu$m are shown in red, {\em Herschel} PACS 100 and 160\,$\mu$m in orange and green respectively, and ALMA 870 
and 1305\,$\mu$m in light and dark blue respectively.  The ALMA sensitivities have been corrected for spatially-resolved emission 
from dust beyond 11.2\,AU (below roughly 20\,K for blackbody grains).  Also shown as a dashed purple line are predictions for the 
EVLA at 45\,GHz in the most favorable configuration (\S3.1).  The filled triangles are evolved, cool white dwarf stage projections 
(\S3.2) of known, dusty A-type stars \citep{su06}, while the progenitor systems are shown as open, inverted triangles in the left panel.  
The observations as a whole were particularly sensitive to evolved main belt analogs, shown as a dotted line in each plot. The star 
symbol in the upper right corner of the left-hand panel is the (undetected) known warm disk, orbiting interior to 0.01\,AU.
\label{fig4}}
\end{figure*}

\subsection{Sensitivity to Dust Emission}

Figure \ref{fig3} details the spectral energy distribution of G29-38, extending from the ultraviolet to submillimeter wavelengths.  
The stellar photosphere is detected from the far-ultraviolet until the near-infrared, where the spatially unresolved emission of the 
inner disk dominates at $\lambda > 2$\,$\mu$m.  Of the suite of available data, the ALMA observations had the best ability
to spatially resolve any emission from dust, but only for relatively cold and distant material beyond 11\,AU.

Following \citet{wya08}, all the available infrared and submillimeter upper limit $F_{\nu}$ at G29-38 were transformed into limiting 
fractional disk luminosities as a function of dust temperature and corresponding orbital radius using 

\begin{equation}
f = 3.4 \times 10^9 F_{\nu} d^2 X_{\lambda} / r^2 B_{\nu}(\lambda,T)
\end{equation}

\noindent
where $d$ is the distance to the star, $r$ the orbital radius of dust grains of temperature $T$, and $B_{\nu}$ is the Planck function.  
The factor $X_{\lambda}$ allows for the faster falloff in emission towards longer wavelengths for non-blackbody -- smaller, warmer 
yet more distant-- grains \citep{wil06}.  These upper limit fractional dust luminosities are plotted as colored curves in Figure \ref{fig4} 
for the {\em Spitzer}, {\em Herschel}, and ALMA observations.  Also included in the plot is a similar curve for the most compact 
configuration of the Expanded Very Large Array (EVLA) at 45\,GHz, with a $1\farcs5$ beam, and a 5.6\,$\mu$Jy continuum 
sensitivity \citep{per11}.

Any blackbody dust grains with temperature below approximately 20\,K would be orbiting beyond 11\,AU and hence be spatially 
resolved to some degree in the ALMA observations, which had a band 7 beam diameter between 1\farcs5 and 1\farcs8, and thus 
capable of resolving ring structures with radii larger than $10.2-12.2$\,AU at 13.6\,pc (see Figure \ref{fig2}).  The plotted ALMA 
sensitivities were corrected for such resolved cases by assuming a face on disk and dividing by the number of beams per ring 
circumference.  The sensitivities of the {\em Herschel} observations were similarly corrected for dust located beyond $50-80$\,AU, 
but it is worth noting that such high fractional luminosities 1) lie above the Figure \ref{fig4} plots and cannot be assumed to be 
optically thin, and 2) would have been detected as spatially resolved emission with ALMA.

\subsection{Projections of Dusty A-type Stars}

Also in Figure \ref{fig4} are plotted model extrapolations for dusty A-type stars from \citet{su06}, as their {\em current} dust properties 
would appear if the host stars were evolved into cool white dwarfs with parameters similar to G29-38.  To begin, the inferred orbital 
radii of the dust components on the main sequence were expanded by a factor $r_{\rm wd}/r_{\rm ms}=M_{\rm ms}/M_{\rm wd}\approx
3.5$, appropriate for G29-38 and a representative value for A stars based on the initial-to-final mass relations (e.g.\ \citealt{wil09,kal08}).  
This change not only leads to a change in dust grain temperatures, but also in their illumination.  The amount of light intercepted by the 
dust decreases by $(r_{\rm ms}/r_{\rm wd})^2$, but the total emitting area of the dust {\em increases} because small grains are not 
removed by radiation pressure \citep{far08}.  On the main sequence the collisional cascade is truncated at a grain diameter that is
approximately $L_{\rm ms}/M_{\rm ms}$ \citep{art88}, and thus the emitting area and fractional luminosity is proportional to $1/\sqrt{
L_{\rm ms}/M_{\rm ms}}$.  

Following \citet{bon10}, considering the realistic emission properties of small grains illuminated by the faint white dwarf, only particles 
larger than 0.1\,$\mu$m contribute significantly to the emission, and thus the change in fractional luminosity will be

\begin{equation}
f_{\rm wd}/f_{\rm ms}\approx(r_{\rm ms}/r_{\rm wd})^2 \times \left( \sqrt{L_{\rm ms}/M_{\rm ms}} / \sqrt{0.1} \right)
\end{equation}

\noindent
Notably, several of the brightest A-star debris disk projections were readily detectable in the ALMA (but not {\em Herschel} or {\em 
Spitzer}) observations, even with the modest number of antennae and corresponding limited sensitivity.  However, it is important 
to note that these modeled points: 1) are based on disks orbiting relatively young, main sequence stars, 2) essentially preserve 
the mass of colliding planetesimals between stellar phases, and 3) allow arbitrarily cold dust, including temperatures below that 
supported by ambient interstellar radiation.  A fixed minimum temperature for dust grains that is substantially warmer than the 3\,K 
cosmic microwave background will favorably impact their detectability with ALMA as such a disk will tend towards the more sensitive 
parts of the Figure \ref{fig4} curves, while the assumption of no further collisional evolution nor dust depletion (e.g.\ wind drag or 
dynamical) over the course of stellar evolution, are admittedly physically unrealistic.  

While small dust grains would be removed during the asymptotic giant phase, they are quickly replenished by collisions.  The collision 
rate of the largest surviving particles can be long, but the smallest grains are repopulated from the integrated collisions of all large bodies, 
resulting in a replenishment timescale equal to their depletion timescale in a steady state \citep{wya11}.  Still, detailed models for the 
post-main sequence evolution of Kuiper belt analogs \citep{bon11,bon10} suggest that detecting these disks in cool white dwarf systems 
like G29-38 is challenging, primarily due to collisional depletion of the disc material.  Such collisional depletion would occur on even shorter 
timescales for asteroid belt analogs favored by the volatile poor abundance patterns seen via atmospheric pollution \citep{gan12,jur12}, 
including G29-38 in particular \citep{xu14}.  And while searches for cold disks have been carried out with {\em Spitzer} in more favorable, 
hot (pre-)white dwarf systems with relatively high stellar luminosity \citep{chu11,su07}, the resulting picture is complicated by binarity and 
dusty outflows associated with their immediate progenitors \citep{cla14}.   

The stochastic planetesimal accretion models of \citet{wya14} place G29-38 among the top 1\% of accretors in terms of incoming
material, but it is also notable among DA stars where instantaneous rates can be inferred from observed metal abundances.  If white
dwarf pollution were correlated with disk brightness on the main sequence, then G29-38 would have evolved from one of the brightest 
main sequence disks and thus imply the highest chance of cold dust detection among polluted white dwarfs.  The fact that it is not 
detected, and the fact that the young dusty A star disks are predicted to be detectable in the absence of depletion of their belts, may 
support, but not strongly, the likelihood that such main sequence disks, on average, become depleted in dust as they progress 
significantly beyond their current, relatively young ages.

These caveats notwithstanding, the analysis and figures indicate that in the absence of significant depletion, or in the case of where 
debris is replenished in these orbital regions (e.g. \citealt{sto14}), that ALMA is an excellent tool to both detect and spatially resolve disks 
at white dwarfs.  Giant impacts or instabilities analogous to the late heavy bombardment would substantially increase the detectability 
of planetesimal disks at white dwarfs, and cold dust can accumulate over long timescales due to the feeble stellar luminosity.  Another 
possibility that favors the retention of substantial disk mass is if the planetesimals have high eccentricity and thus a reduced collision 
rate \citep{wya10}.  While the average planetesimal disk at white dwarfs may be depleted, some fraction of the population may remain 
salient.  The final ALMA array of 66 dishes should more than double the sensitivity, all else being equal, and thus be capable of detecting 
circumstellar dust more representative of current A stars, as well as asteroid belt analogs, now cold and expanded beyond several AU.

\begin{figure}
\includegraphics[width=84mm]{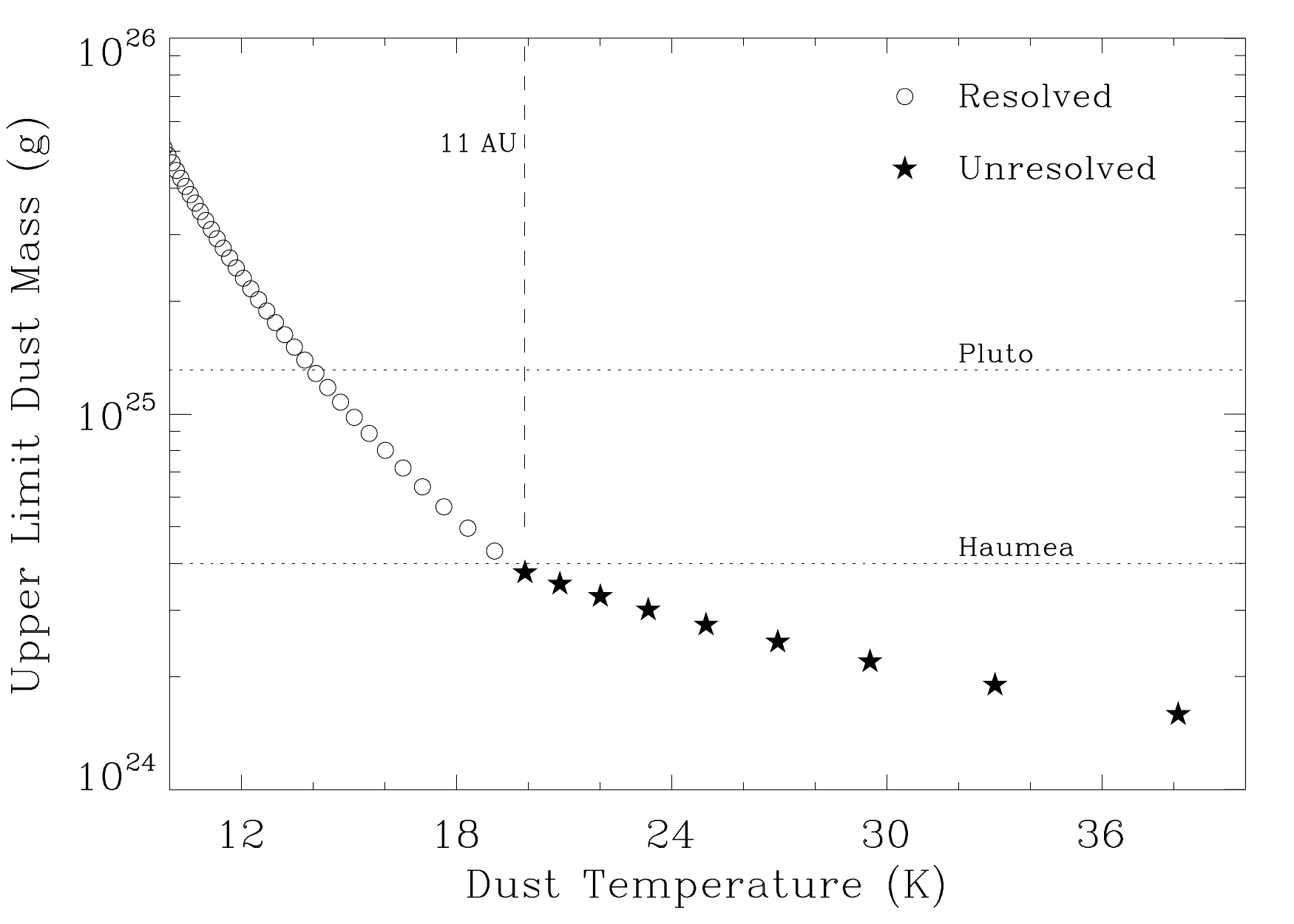}
\caption{Dust mass limits calculated from the achieved ALMA upper limit at 870\,$\mu$m of 0.17\,mJy.  For dust warmer than around
20\,K the emission is expected to remain spatially unresolved in the observations, while cooler dust would have been spread across
many beams and thus the sensitivity drops accordingly.
\label{fig5}}
\end{figure}
\subsection{Dust Mass Limits}

To derive upper limits on the mass of dust present around G29-38, the 870\,$\mu$m observations were the most sensitive for realistic 
temperatures.  Emission at this wavelength traces predominantly millimeter sized dust particles and is most likely optically thin. In the 
absence of any detailed information on the composition and size of any grains associated with this source, the average value of dust
opacity is adopted here, $\kappa=1.7$\,cm$^2$\,g$^{-1}$ at 870\,$\mu$m, noting that the actual opacity may be anywhere in the range 
$0.2-4.0$\,cm$^2$\,g$^{-1}$ \citep{dra06}.

Using a radial dependence of the dust temperature identical to that derived in \S3.1 and plotted in Figure \ref{fig4}, the adopted upper
limit flux of $F_\nu=0.17$\,mJy/beam is converted to the dust mass using 

\begin{equation}
m(T)=F_\nu d^2/\kappa B_\nu(T)
\end{equation}

\noindent
A range of radial distances $r$ were explored, where dust emission falls within one half of the primary beam of a 12\,m ALMA 
antenna at 870\,$\mu$m, or up to 60\,AU from the star.  Having no information on the radial or azimuthal distribution of dust, nor its 
projected distance in the plane of the sky, the calculations assume a simple, circular, and face-on ring of radial thickness much less 
than the beam size (22\,AU at 13.6\,pc).
 
Spatially unresolved emission allows direct conversion from the observed flux to the mass of the dust using the above expression, while 
at $2r>22$\,AU the ring would become resolved and the observed flux per beam is then a fraction of the total flux in the ring.  Upper limits 
derived in this way are shown in Figure \ref{fig5}, where only high dust masses greater than that of Pluto ($1.3\times10^{25}$\,g) can be 
ruled out beyond 11\,AU in the spatially resolved regime.  In the case of unresolved emission within 11\,AU, the upper limit dust masses 
are relatively low and comparable to a few times the mass of Ceres ($9.4\times10^{23}$\,g).  

Coincidentally, the unresolved dust mass limits fall within a factor of a few of the highest known masses of metals residing in the outer 
layers of polluted white dwarfs with significant convection zones \citep{duf10,far10a}.  However, there is no reason to expect a physical 
connection between collisionally generated dust masses in outer planetesimal belts (which requires replenishment over appropriate 
timescales), and the amount of mass that is transported via (presumably) intact parent bodies to the tidal disruption radius, and later 
the stellar surface, of polluted white dwarfs.

\section{OUTLOOK}

Debris from the inferred parent population of planetesimals at G29-38 remains undetected with {\em Herschel} PACS at 100 and 
160$\mu$m, and ALMA in early and relatively shallow observations at 870 and 1305\,$\mu$m.  While the ALMA observations were
best suited to outer regions analogous to evolved Kuiper belt analogs, the {\em Herschel} data were uniquely sensitive to an evolved,
asteroid-like belt at G29-38 and in general to dust at temperatures and orbital regions intermediate to the relatively warm dust seen
at polluted white dwarfs and the cooler dust often detected at main sequence stars.  The non-detection at G29-38 is not wholly 
unexpected, as disk evolution models supported by observations of main sequence stars predict that the available mass in both 
dust and parent bodies decreases significantly over timescales of several hundred Myr \citep{wya07}, and this depletion is likely 
enhanced during the post-main sequence \citep{bon10}.  

These data collectively preclude a relatively bright, $L_{\rm IR}/L_*>10^{-4}$ disk in the general vicinity of 10\,AU around the nearest
and brightest polluted white dwarf with an infrared excess.  From orbital expansion alone, a planetesimal belt currently in within 10\,AU
would have orbited within the terrestrial zone -- interior to the water ice line -- during the main sequence, and thus be consistent with
the suspected source regions for parent bodies of the disk of disrupted and polluting material in G29-38 and a growing number of 
white dwarfs where detailed abundance measurements allow a robust assessment of their volatile content \citep{koe14,xu14,far13,
gan12,jur12}.  Therefore, any dusty asteroid-like belt must lie below these detection limits. 

However, recent work suggests that it is difficult for sufficient material to survive to the white dwarf phase in this inner region 
\citep{fre14,deb12}.  A parent body origin in an outer, Kuiper-like belt remains consistent with these long wavelength observations 
\citep{bon11}, but the volatile deficiency combined with substantial parent body masses, both inferred via atmospheric pollution 
remains to be explained.

While significant uncertainty remains, ALMA is the only current facility able to empirically constrain the source regions for the parent
bodies of the planetary debris surrounding and falling onto white dwarfs like G29-38.

\section*{ACKNOWLEDGMENTS}

The authors are grateful to Anita Richards and the UK ARC at the University of Manchester for observation and data analysis 
support, and to the anonymous reviewer for suggestions that improved the manuscript.  This paper makes use of the following ALMA 
data: ADS/JAO.ALMA \#2011.0.00851.S (PI Farihi).  ALMA is a partnership of ESO (representing its member states), NSF (USA), and 
NINS (Japan), together with NRC (Canada), NSC and ASIAA (Taiwan), in cooperation with the Republic of Chile.  The Joint ALMA 
Observatory is operated by ESO, AUI/NRAO and NAOJ.  {\em Herschel} is an ESA space observatory with science instruments 
provided by European-led Principal Investigator consortia and with important participation from NASA. J. Farihi gratefully acknowledges 
the support of the STFC via an Ernest Rutherford Fellowship.  A. Bonsor acknowledges the support of the ANR-2010 BLAN-0505-01 
(EXOZODI).  M. C. Wyatt and O. Pani\'c are grateful for the support of the European Union through ERC grant number 279973.

\label{lastpage}

\end{document}